\documentclass{PoS}
\newcommand{\eg}{\textit{e.g.}} 
\newcommand{\ie}{\textit{i.e.}} 
\def\Journal#1#2#3#4{{#1}\,{#2}, #3 (#4);} 
\newcommand{\etal}{et al.}
\newcommand{\ApJ}{Astrophys. J.}

\newcommand{\PRL}{Phys. Rev. Lett.}
\newcommand{\PRD}{Phys. Rev. D}

\newcommand{\ASR}{Adv. Space Res.}

\newcommand{\JGR}{J. Geophys. Res.}

\title{New results in solar modulation modeling\\ in light of recent cosmic-ray data from space}
\ShortTitle{Numerical Models for Solar Modulation of GCRs}

\author{{Bruna Bertucci}\thanks{E-mail: {bruna.bertucci@pg.infn.it}}\\
Department of Physics and Earth's Science, Universit{\`a} di Perugia, and INFN - Perugia, I-06100 Perugia, Italy\\
}

\author{{Emanuele Fiandrini}\thanks{E-mail: {emanuele.fiandrini@pg.infn.it}}\\
Department of Physics and Earth's Science, Universit{\`a} di Perugia, and INFN - Perugia, I-06100 Perugia, Italy\\
}

\author{{Behrouz Khiali}\thanks{E-mail: {behrouz.khiali@ssdc.asi.it}}\\
INFN - Roma Tor Vergata, and Space Science Data Center - ASI, I-01100 Roma, Italy\\
}

\author{{Nicola Tomassetti}\thanks{E-mail: {nicola.tomassetti@pg.infn.it}}\\
Department of Physics and Earth's Science, Universit{\`a} di Perugia, and INFN - Perugia, I-06100 Perugia, Italy\\
}

\abstract{
Thanks to space-borne experiments such as the AMS-02 and PAMELA missions in low-Earth orbit,
along with the Voyager spacecrafts in the interstellar space, a large collection of multi-channel
and time-resolved Galactic cosmic ray (GCR) data has recently become available.
Here we present an improved measured-validated model of the \emph{solar modulation} effect, \ie, 
the temporal evolution of the GCR flux inside the heliosphere caused by the 11-year variability cycle of the Sun's magnetic activity.
We present our improved modeling of the structure of the heliosphere,
the physical mechanisms of diffusion, drift, and energy losses of GCR particles in the heliosphere.
We present our results for the temporal dependence of the key model parameters and their relationship with solar activity proxies.
We discuss implications for the GCR transport in magnetic turbulence, and new insights on our understanding of the solar modulation phenomenon.
} 
\FullConference{36th International Cosmic Ray Conference - ICRC2019 -\\
		July 24th - August 1st, 2019\\
		Madison, WI, U.S.A.}

\begin{document}

\section{Introduction}      
\label{Sec::Introduction}   

The spectrum of Galactic cosmic rays (GCRs) observed near-Earth is significantly different from the
so-called Local Interstellar Spectrum (LIS), \ie, their spectrum  beyond the boundaries of the heliosphere,
because GCRs are subjected to several physical mechanisms during their propagation through the interplanetary space.  
More specifically, charged particles in the heliospheric plasma can be spatially diffused, magnetically drifted,
advected and decelerated by the outflowing solar wind and its embedded magnetic field. 
The modifications of the GCR intensities and energy spectra are known to vary with time, in connection with the Sun's magnetic activity.
This phenomenon is referred to as \emph{solar modulation} of GCR, and it is observed to change periodically, following the
periodical change in the sunspot number (SSN) observed in the solar corona.
The SSN is a widely used in solar physics, \eg, as proxy for solar activity or characterizing the Solar Cycle,
as well as for studying its correlation with the GCR modulation properties \cite{Usoskin1998,Ross2019}.

A thorough understanding of solar modulation is then important, in the physics of GCRs,
either to infer their LIS's or to investigate the dynamics of charged particles in the heliospheric turbulence. 
Understanding the evolution of the GCR radiation is also essential for astronauts and the electronic components
radiation hazard during long-duration missions.
Along with the Voyager-1 data beyond the heliosphere \cite{Cummings2016},
the new precise data from AMS-02 \cite{Aguilar2018PHeVSTime,Aguilar2018LeptonVSTime} and PAMELA \cite{Adriani2013,Martucci2018}
experiments offer a unique possibility to study the solar modulation over a long period of time.

\section{The Numerical Model}  
\label{Sec::Model}             

\begin{figure*}[hbt]
\centering
\includegraphics[clip,trim=1cm 0.1cm 0.01cm 0.1cm, width=0.52\textwidth]{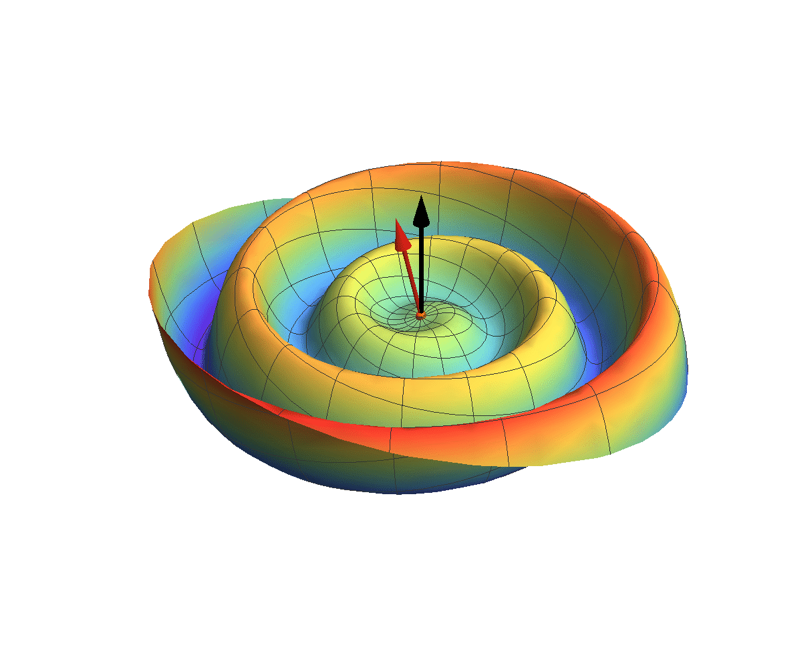} 
\quad
\includegraphics[width=0.40\textwidth]{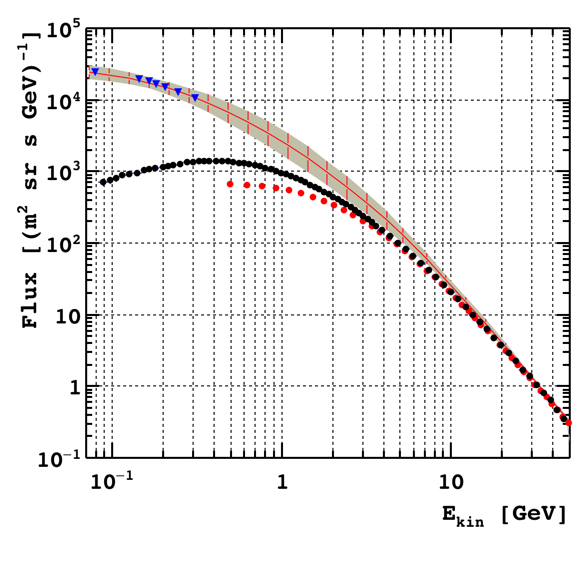}
\caption{Left: Sketch of the \emph{heliospheric current sheet}, the thin surface where magnetic polarity changes from north to south.
  The waviness of this surface is related by the \emph{tilt angle}, shown in the figure.
  Right: GCR flux calculations for LIS of GCR protons (solid line)
  and its uncertainties (gray band). The data are from Voyager-1 (blue triangles), AMS-02 (red dots) and PAMELA (black dots).
}
\label{Fig::ccHCS} \label{Fig::ccProtonLIS} 
\end{figure*}

The propagation of GCRs in the heliosphere is described by the Parker-Krymsky equation \cite{Parker1965}:
\begin{equation}
\label{Eq::Parker}
\frac{\partial f}{\partial t}
=  \nabla\cdot [\mathbf{K}^{S}\cdot\nabla f ]
- (\vec{V}_{sw} + \vec{V}_D) \cdot\nabla f 
+ \frac{1}{3}(\nabla \cdot\vec{V})\frac{\partial f}{\partial (ln R)}  
\end{equation}
The equation describes the time evolution of the GCR phase space density $f(t,R)$,
where $R=p/Z$ is the GCR particle rigidity (momentum/charge ratio),
$\vec{V}_{sw}$ is the solar wind speed, $\vec{V}_{D}$ is the guiding center drift speed,
and $\mathbf{K}^{S}$ is the symmetric part of the GCR diffusion tensor \cite{Potgieter2013,Moraal2013}. 
The GCR flux {\it J=J(t,R)} is related to $f$ by $J = R^{2}f$.
In this work, the equation is solved in steady-state conditions ($\partial/\partial{t}=0$),
using the stochastic approach \cite{Strauss2017}, 
based an upgraded and customized version of the code \emph{Solarprop} \cite{Kappl2016,Tomassetti2017BCUnc}. 
We use a 2D model, with radius $r$ and heliolatitude $\theta$ as spatial coordinates.
The heliosphere is regarded as a spherical cavity, placed around the Sun, where the wind flows radially.
We place the termination shock (TS) at $r_{\rm TS}=85$\,AU from the Sun. Here the wind drops to subsonic speed.
The heliospheric boundary is taken at $r_{\rm HP}=122$\,AU, the heliopause (HP), where
the wind vanishes and the GCR flux joins its interstellar value.
Within this structure, the Earth lies in the equatorial plane at $r_{\rm E}=$1\,AU.
Along with its radial direction, the speed of the solar wind is known to depend upon latitudinal and solar activity.
The latitudinal dependence appears well pronounced during period of minimum activity (low SSN) while, in contrast,
the wind profile becomes highly symmetric during solar maximum (high SSN).
Furthermore, in the equatorial plane the speed is highly stable with time, so that it is often set to the constant value of $400$\,km\,s$^{-1}$.
In our model, we adopt a global parameterization for $V_{sw}(r,\theta)$, as in \cite{Potgieter2014}.
Regarding its influence on the GCR modulation, the wind transports a frozen-in turbulent
Heliospheric Magnetic Field (HMF), wounded up in a rotating spiral structure. 
Of great importance for GCR modulation is the Heliospheric Current Sheet (HCS), sketched in Fig.\,\ref{Fig::ccHCS}
The HCS is a co-rotating magnetic structure which divides the HMF into two hemispheres of opposite polarity.
%
%
%
The angular extension of the HCS is described by the so-called \emph{tilt angle} $\alpha$.
The tilt angle is the angle between the Sun's rotational axis and its main magnetic axis. 
It sets the level of the HCS waviness, which is time dependent.
The $\alpha$-angle is constantly monitored by the  Wilcox Solar Observatory (WSO) on a 10-day basis \cite{Hoeksema1995,Alanko2007}.
It ranges from a few degrees during solar minimum (flat HCS) to nearly 90$^{0}$ during maximum and reversal (wavy HCS).
In particular, magnetic drift of GCRs is important across the wavy layer of the HCS.
More in general, the motion of GCR particles in the HMF is decomposed in a regular gradient-curvature large-scale motion
and a HCS drift motion, both occurring on the background average HMF, plus an erratic random-walk scattering off the
small-scale HMF turbulence, from which GCR diffusion originates.
All these effects can be included in a diffusion tensor $\mathbf{K}$, made of a symmetric part that describes
the diffusion and a asymmetric part that describes drift:
$\mathbf{K}=\mathbf{K}^S+\mathbf{K}^A$, with $K_{ij}^S = K_{ji}^S$ and $K_{ij}^A = -K_{ji}^A$.
The elements of the symmetric tensor that describes the diffusion $K^S_{ij}$ can be derived from kinetic theory,
following a microscopic approach from GCR pitch-angle scattering off the random HMF irregularities.
A key input parameter for the cosmic ray transport is the parallel spatial diffusion coefficient $K_{\parallel}$, which is
usually expressed in terms of the mean free path $\lambda_{\parallel}$ along the background HMF:
as $K_{\parallel} = \beta c \lambda_{\parallel}/3$,  where $\beta=v/c$ is the GCR speed.
GCRs are also subjected to perpendicular diffusion, with an associated length $\lambda_{\perp}$,
that is usually assumed to be proportional to the parallel diffusion length:
$\lambda_{\perp}= \xi \lambda_{\parallel}$, with  $\xi \cong\,0.02$ \cite{Giacalone1999}.
The scattering of the GCR on the HMF irregularities is maximum when the GCR Larmor radius $r_{L}$ 
is of the same order as the spatial scale size of the irregularities  $\hat{\lambda}$.
Thus, the scattering for GCR particles with rigidity $R$ will depends upon the HMF power spectrum
around the wave number $k$, corresponding to the spatial scale $\sim r_{L}(R)$.
The essential dependence of $\lambda_{\parallel}$  on the HMF power spectrum can be therefore expressed as:
\begin{equation}
\lambda_{\parallel} \sim r^{2}_{L}\frac{\langle B^{2}\rangle}{w(k_{\rm res})} \sim R^{2}/w(k_{\rm res})
\label{Eq::Resonance}
\end{equation}
where $\langle B^{2}\rangle$ is the mean square value of the background field and $w(k_{\rm{res}})$ is the power spectral density of the HMF fluctuations.
The spatial scale of the resonant component of the spectrum is $\hat{\lambda} = 2\pi/k_{\rm{res}}$.
From the cyclotron resonance condition, $r_{L} \sim \hat{\lambda}$, particles with rigidity $R$ will resonate at $k_{\rm{res}} \sim 1/R$. 
The $w(k)$ follows a power-law $w(k) \propto k^{-\eta}$, where the index $\eta$ depends on the type and on the spatial scales
of the turbulence energy cascade \cite{Kiyani2015}. 
Thus, $\lambda_{\parallel}$ depends on the turbulence spectral index as $\lambda_{\parallel} \sim R^{2-\eta}$.
On a wide range of scale,  at least three regimes (with different index $\eta$)
can be distinguished for the HMF power spectrum \cite{Kiyani2015} 
that, in turn, will correspond to three diffusion regimes for GCRs:
the \emph{$1/f$ range} at large scales (\ie, small $k$, high rigidity),
the \emph{inertial range} at intermediate scales,
and \emph{sub-ion range} at the smallest scales (large $k$, low rigidity).
At rigidity above few hundreds MVs, relevant to this work, GCRs resonate in the $1/f$ and $inertial$ scale range.
Thus, a good parameterization for the rigidity dependence of $\lambda_{\parallel}$
is the \emph{double power-law} function, defined by two indices and a critical rigidity value.
In this work, we have adopted the following parametric description for the parallel diffusion coefficient $K_{\parallel}$ \cite{Potgieter2014}:
\begin{equation}\label{Eq::Par_diff}
K_{\parallel} = \frac{K_0}{3}\beta \left(\frac{B_0}{B}\right)   \left(\frac{R_0}{R}\right)^a 
 \times \left[ \frac{(R/R_0)^h + (R_k/R_0)^h }{1 + (R_k/R_0)^h} \right]^{\frac{b-a}{h}} 
\end{equation}
where $K_{0}$ is a constant in units of $10^{23}$ $cm^2 s^{-1}$,  with $R_{0}\equiv$\,1\,GV,
$B$ the HMF magnitude and $B_{0}$ the \emph{local} field value (at 1 AU). 
The parameters $a$ and $b$ are two power indices that set the slope of the rigidity dependence, respectively,
below and above the rigidity $R_{k}$. The parameter $h$ determines the smoothness of the transition between the two regimes.
The perpendicular diffusion follows from $\lambda_{\perp}= \xi \lambda_{\parallel}$,
with the addition of polar corrections (based on the Ulysses data at high latitudes) made to
$\lambda_{\perp}$ and to the HMF model  \cite{Simpson1996,Heber1998}.
The parameter set ${K_{0}, a, b}$ then specifies the diffusion of GCRs on the HMF turbulence.
Because these parameters depend on the properties of the background plasma,
they may be subjected to temporal evolution following the Solar Cycle \cite{Manuel2014}.
To test the time evolution of these parameters, we use the recent data on GCR protons from 
AMS-02 and PAMELA \cite{Aguilar2018PHeVSTime,Adriani2013,Martucci2018}.

At this point it is important to note that
Eq.\,\ref{Eq::Parker} requires the flux at the HP to be specified as boundary condition,
for a given GCR particle species.
Determining the GCR LIS requires the development of dedicated models of Galactic physics processes, \eg, 
from the distribution of GCR sources, particle acceleration mechnisms and diffusive propagation in the Galactic
turbulence \cite{Grenier2015}.
In the work presented here, the GCR proton LIS relies on improved 
calculations from recent works \cite{Tomassetti2015TwoHalo,Feng2016,Tomassetti2018PHeVSTime}.
In these calculations, the proton LIS was tightly constrained with low-energy data (at $\sim$\,100\,-500\,MeV of kinetic energy)
collected from Voyager-1 beyond the HP \cite{Cummings2016}
and AMS-02 precision measurement performed in the high-energy regime ( $E\gtrsim$\,100\,GeV), \ie,
where solar modulation effects become asymptotically negligible \cite{Aguilar2018PHeVSTime,Aguilar2015Proton,Aguilar2015Helium}.
The GCR proton LIS of this work agrees fairly well with other spectra proposed recently
\cite{Boschini2017,Corti2016,Corti2019,Tomassetti2017TimeLag,Tomassetti2015PHeAnomaly,Tomassetti2017Universality}. 
Our LIS is shown in Fig.\,\ref{Fig::ccProtonLIS}, where comparison is made with the data from Voyager-1, AMS-02, and PAMELA.

\section{The parameter extraction}   
\label{Sec::Model}                   

Our model has six free parameters, in principle time-dependent, 
that have to be determined using GCR or Solar data at a given epoch $t$: 
the tilt angle of the HCS, $\alpha(t)$, the HMF intensity at Earth's location $B_{0}(t)$, the HMF polarity $A(t)$,
the global normalization factor of the diffusion tensor $K_{0}(t)$,
and  two spectral indices of its rigidity dependence, $a(t)$ and $b(t)$,
defined at rigidity below and above the break parameter $R_{k}$.
In this work we follow the quasi steady-state approach,
where the time-dependence of the problem is described as a continuous series of steady-state solution of Eq.\,\ref{Eq::Parker}.
In practice, time-series of data are used to extract a corresponding time-series of best-fit parameters.
For a given epoch defined by the time $t$, the parameters $\alpha(t)$, $B_{0}(t)$ and $A(t)$ can be directly determined
with Solar observations from WSO. These parameters describe the average status of the HMF in a given epoch.
Since $\alpha$ and $B_{0}$ are subjected to continuous variations, we have computed an average value at
a given epochs $t$, performed over a time window  $[t-\Delta{T}, t]$, with $\Delta{T}=$\,12\,months.
The window is chosen so that the average values of $\hat{\alpha}$ and $\hat{B}_{0}$ reflect
the average HMF conditions sampled by GCRs when propagating from the heliopause to Earth \cite{Tomassetti2017TimeLag}. 
In fact, the model assumes a static heliosphere, where its status is frozen during the travel of GCRs toward Earths.

The remaining parameters $K_{0}$, $a$, and $b$ are those describing the properties of GCR diffusion
on the heliospheric plasma. They have been determined using GCR proton measurements from AMS-02 and PAMELA.
In practice, to fit the GCR data, we have built a six-dimensional discrete grid of the model parameters
vector $\vec{q}=$ ($\alpha$, $B_0$, $A$, $K_0$,  $a$, $b$), for about $10^{6}$ grid nodes, \ie, parameter configurations.
For each configuration, a differential energy spectrum of GCR protons $J_{m}(E, \vec{q})$ was calculated.
Each calculation consist in the Monte-Carlo simulation of about 6$\times\,10^{4}$ pseudo-particles,
backwardly-propagated from Earth to the HP, and then re-weighted for their LIS, as to obtain the locally modulated GCR flux.
In practice, $2\times 10^3$ pseudo-particles were simulated for each of the 60 energy bins,
ranging from 20 MeV to 200 GeV and following a logaritmically increasing step.
Once the proton grid has established, the time-resolved GCR proton data has been utilized.
From the data at a given epoch (time $t$), the corresponding quantities $A$, $\hat{\alpha}$, and $\hat{B_{0}}$ are determined in the first place. 
Then, for each GCR data set $J_{d}(E,t)$, a $\chi^{2}$ statistics was computed for all the modulated fluxes in the model array:
\begin{equation} \label{Eq::ChiSquare}
  \chi^{2}(\vec{q})= \sum_{i}  \frac{\left[ J_{d}(E_{i},t) - J_{m}(E_{i}, \vec{q}) \right]^{2}}{\sigma^{2}(E_{i},t)}
\end{equation} 
the model $J_{m}$ energy flux at a given energy was computed by interpolated along on the grid,
so that the resulting $\chi^{2}(\vec{q})$ function of Eq.\,\ref{Eq::ChiSquare} is a continuous function of the parameters
(with the exception of the polarity parameter $A$, that takes two values).
The errors are given by $\sigma^{2}(E_{i},t) = \sigma_{d}^{2}(E_{i},t) + \sigma_{mod}^{2}(E_{i},t)$, \ie,
by the sum in quadrature of two contributions:  errors in the experimental measurements, for the $i$-th energy bin,
and theoretical uncertainties in the calculated flux $J_{m}$, modulated, related to LIS and to the Monte-Carlo generation statistics.
Physically, this is due to adiabatic losses, that prevent a large fraction of low energy to reach the Earth.
To reduce fluctuations, a high number of pseudo-particle trajectories has to be simulated.
In particular, we generated about $2\times10^{3}$ pseudo-particles per energy bin per run,
which demanded a considerable CPU time. 
The evaluation of the flux fluctuations was done as follows.
The ratio between the modulated flux and the corresponding LIS  $J_{mod}/J_{\rm{LIS}}$, is approximately equal to the ratio $\sim N_{m}/N_{G}$,
where $N_{m}$ is the number of pseudo-particles that reach the HP with energy $E$, and $N_{G}$ is the number of pseudo-particles generated at the same energy.
Since the propagation process is stochastic, we expect that the relative error of the modulated flux decreases
as ${\delta}J_{mod}/J_{mod} = 1/\sqrt{N_{m}}$ with $N_{m} = \left(J_{mod}/J_{LIS}\right)N_{G}$.
The resulting errors amount to $\sim$\,10$\div$20\,\% for GCR energy of 20 MeV,
and become constant at the level of $\sim\,{2\%}$ at GCR energy above a few GeVs.
The determination of the remaining GCR diffusion parameters ($K_{0}(t)$, $a(t)$ and $b(t)$) proceeds as follows.
Given a data set $J_{d}(E,t)$, for every parameter $q$ = $K_{0}(t)$, $a(t)$ and $b(t)$, the corresponding $\chi^{2}(q)$ distribution
is evaluated, its absolute minimum $\chi^{2}$ is found, and the corresponding parameter $q_{\rm best}$ is then taken as best-fit.
To find the minimum two methods were implemented, both based on a multidimensional grid interpolation,
and the associated systematic errors were accounted for.

\section{Results and discussion}  
\label{Sec::Results}              

\begin{figure*}[hbt]
\centering
\includegraphics[clip,trim=1cm 0.1cm 0.01cm 0.1cm, width=0.90\textwidth]{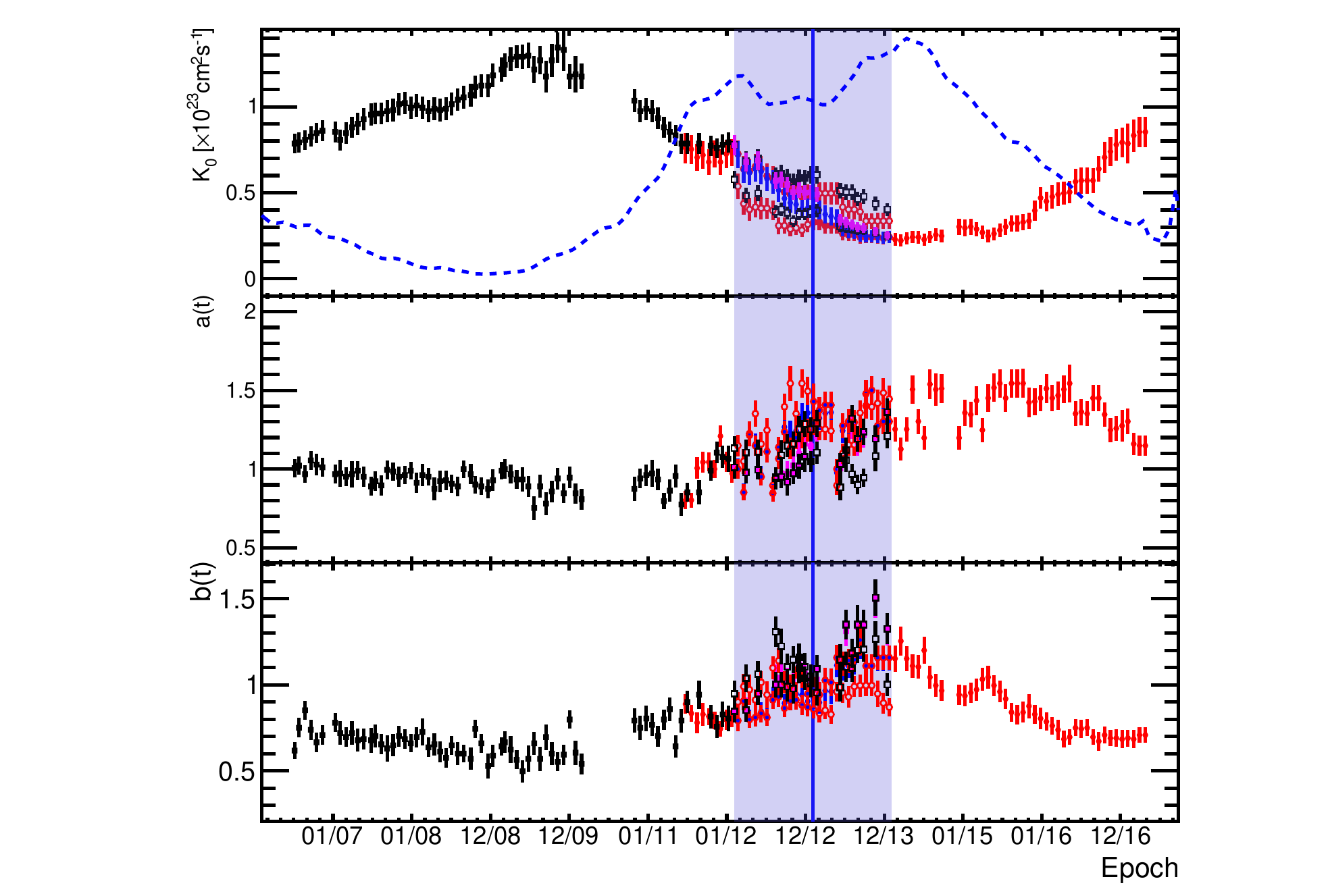}
\caption{Time series of the best-fit model parameters $K_{0}$, (top panel) spectral index $a$ (middle), and spectral index $b$ (bottom).
  The temporal dependence of the smoothed SSN are shown as dotted line in the top panel, renormalized in order to fit in the plot.
  The epoch of HMF polarity inversion $T_{inv}$ is shown as vertical blue line; the reversal period as shaded band \cite{Sun2015}.
  Best-fit result obtained with PAMELA (black) and AMS-02 (red) data are shown.
  Inside the reversal band, the results are shown under both polarities $A<0$ and $A>0$,
  as a smooth transition is expected between the two values.
  The modeled transition across the reversal is also displayed, as blue solid line for AMS-02 and magenta for PAMELA.}
\label{Fig::BestFitParametersVSTime} 
\end{figure*}

For each free parameter, our modified least-square minimization procedure returns a time-series of
its best-fit values and their uncertainties. The results of the fitting are summarized in Fig.\,\ref{Fig::BestFitParametersVSTime},
showing the diffusion scaling $K_{0}$ (top panel) and its spectral indices $a$ (middle) and $b$ (bottom). 
The colors represent the data sets used to perform the fits, \ie, time-resolved proton measurements from AMS-02 (red full dots) and PAMELA (black full squares).
It is important to note that the period covered by our study spans across a significant fraction of Solar Cycle,
including the interesting phase of magnetic reversal, \ie, when the HMF polarity inversion takes place.
We indicate by $T_{inv}$ the epochs of reversal, occurred between 2012 and 2013, which is shown in the figure
by vertical blue line, along with its uncertainty band \cite{Sun2015}.
We can consider three relevant epochs:
(i) $t \ll T_{\rm rev}$, before HMF reversal and with negative polarity ($A<$0),
(ii) $t \gg T_{\rm rev}$, after reversal and under positive polarity ($A>$0),
(iii) $t \sim T_{\rm rev}$ during the reversal, when the HMF is disorganized so that its polarity $A$ is not well defined.
In the top panel of Fig.\,\ref{Fig::BestFitParametersVSTime}, the smoothed SSN is also shown,
as dotted blue line, renormalized as to fit the scale. The SSN is a good proxy for solar activity
and for studying its correlation with GCR modulation properties \cite{Ross2019}.
From the figure, it can be seen that the diffusion normalization parameter $K_{0}$ evolves with time, following a well defined pattern. 
As clear from the figure, the change in $K_{0}$ appears well anti-correlated with the smoothed SSN, \ie, with solar activity.
In particular, GCR diffusion is faster during solar minimum, and slower during maximum.
This behavior is at the basis of the time-dependent modulation phenomenon.
In fact, a change in diffusion produces a higher (lower) intensity during epochs of solar minimum (maximum) and vice-versa. 
In the ideal limit of infinitely fast diffusion, the modulation effect would disappear  (\eg, without Sun, or in the limit of high-energy GCRs).
During the HMF reversal, it can also be seen that the best-fit $K_{0}$ curve has a little discontinuity,
following the discontinuity of the HMF polarity that switches from $A=-1$ to $A=+1$, across $t=T_{\rm rev}$.
In practice, magnetic reversal occurs gradually, across the region of the shaded band.
In fact it involves the magnetic flip of the north/south poles and the ri-organization of the HMF.
This process has a duration of about one year, as suggested by HMF observations \cite{Sun2015}
and found by AMS-02 data on the $e^{+}/e^{-}$ ratio evolution \cite{Aguilar2018LeptonVSTime}.
For the diffusion parameter values, one may expect a similar transition occurring during this phase.
In Fig.\,\ref{Fig::BestFitParametersVSTime}, the best-fit $K_{0}$-values are shown under both polarities in the reversal phase.
A smooth transition between the two fits is also shown.

Interestingly, we also found that the diffusion spectral indices $a$ and $b$ show clear temporal dependence.
In our work, the index $a$ is related to the power spectrum in the $1/f$ \emph{range}, while $b$ is related
to the \emph{inertial range} of the turbulent energy cascade of HMF.
In both ranges, our parameters are in agreement with the measured slopes of the HMF power spectrum on Jan-Feb 2007 \cite{Kiyani2015}.
In most of numerical models of solar modulation, these parameters are usually assumed to be time-independent.
Variations in these parameters imply unexpected changes in the HMF turbulence spectrum. This is also suggested
in other recent works \cite{Usoskin2019,Borovsky2012,Horbury2005}.

\section{Acknowledgements}  
\label{Sec::Results}        

We acknowledge the support of ASI
under agreement \emph{ASI-UniPG 2019-2-HH.0}.
In this work, we made use of
the \emph{ASI-SSDC Cosmic-Ray Database} at ASI,
the \emph{SILSO/SIDC} SSN Database at the Royal Observatory of Belgium,
and real-time data from the \emph{Wilkox Solar Observatory}.
We thank Todd Hoeksema for support with the WSO solar data data.



\begin{thebibliography}{99} 

\bibitem{Usoskin1998}
  Usoskin, I.G., \etal, 
  \href{https://doi.org/10.1029/97JA03782}{\JGR{} 103, 9567 (1998)}
  
\bibitem{Ross2019}
  Ross, E., Chaplin, W.,
  \href{https://doi.org/10.1007/s11207-019-1397-7}{Sol. Phys. 294, 8 (2019)}
  
 \bibitem{Cummings2016} 
   Cummings, A. C., \etal,
   \href{http://dx.doi.org/10.3847/0004-637X/831/1/18}{\ApJ{} 831, 18 (2016)} 
      
\bibitem{Aguilar2018PHeVSTime}
  Aguilar, M., \etal, 
  \href{https://journals.aps.org/prl/abstract/10.1103/PhysRevLett.121.051101}{\PRL{} 121, 051101 (2018)}

\bibitem{Aguilar2018LeptonVSTime}
  Aguilar, M., \etal, 
\href{https://journals.aps.org/prl/abstract/10.1103/PhysRevLett.121.051102}{\PRL{} 121, 0511012 (2018)}

\bibitem{Adriani2013}
  Adriani, O., \etal,
  \href{https://iopscience.iop.org/article/10.1088/0004-637X/765/2/91/pdf}{\ApJ{} 765, 91, (2013)}
  
\bibitem{Martucci2018}
  Martucci, M., \etal, 
  \href{http://dx.doi.org/10.3847/2041-8213/aaa9b2}{\ApJ{} 854, L2 (2018)}

\bibitem{Parker1965}
  Parker, E. N.,
  \href{http://dx.doi.org/10.1016/0032-0633(65)90131-5}{Planet. Space Sci. 13, 9 (1965)} 
  Krymsky, G., 
  Geomagn. A\'{e}ron. 4, 977-987 (1964)
  
\bibitem{Potgieter2013}
  Potgieter, M. S.,
  \href{http://dx.doi.org/10.12942/lrsp-2013-3}{Living Rev. Solar Phys., 10, 3 (2013)}

\bibitem{Moraal2013}
  Moraal, H.,
\href{https://dx.doi.org/10.1007/s11214-011-9819-3}{Space Sci Rev. 176 299 (2013)}
  
\bibitem{Strauss2017}
  Strauss, R. D., Effenberger, F.,
  \href{http://dx.doi.org/10.1007/s11214-017-0351-y}{Space Sci Rev (2017)}

\bibitem{Kappl2016}
  Kappl, R.,
  \href{https://doi.org/10.1016/j.cpc.2016.05.025}{Comp. Phys. Comm. 207 (2016) 386-399}

\bibitem{Tomassetti2017BCUnc}
  Tomassetti, N.,
  \href{http://dx.doi.org/10.1103/PhysRevD.96.103005}{\PRD{} 96, 103005 (2017)}
  
\bibitem{Potgieter2014}
  Potgieter, M. S., \etal, 
  \href{http://dx.doi.org/10.1007/s11207-013-0324-6}{Sol. Phys. 289, 391-406 (2014)} 

\bibitem{Hoeksema1995}
  Hoeksema, J. T.,
  \href{https://doi.org/10.1007/BF00768770}{Space Sci. Rev. 72, 137-148 (1995)}

\bibitem{Alanko2007}
  Alanko-Huotari, K., Usoskin, I. G., Mursula, K. Kovaltsov, G. A.,
  \href{https://doi.org/10.1016/j.asr.2007.02.007}{\ASR{} 40, 1064-1069 (2007)};
  
\bibitem{Giacalone1999}
  Giacalone, J., \& Jokipii, J. R., 
  \href{https://doi.org/10.1086/307452}{\ApJ{} 520, 204 (1999)}

\bibitem{Kiyani2015}
  Kiyani K. H., Osman K. T., Chapman S.C., 
  \href{http://dx.doi.org/10.1098/rsta.2014.0155}{Phil. Trans. R. Soc. A 373: 20140155 (2015)} %
  
 \bibitem{Simpson1996}
   Simpson, J. A.,
   \href{http://dx.doi.org/10.1007/BF02508134}{Nuovo Cimento C, 19C, 935-943 (1996)}

 \bibitem{Heber1998}
   Heber, B., \etal, 
   \href{http://dx.doi.org/10.1029/97JA01984}{J. Geophys. Res., 103, 4809 (1998)}

\bibitem{Manuel2014}
  Manuel, R., Ferreira, S. E. S., Potgieter, M. S.,
  \href{http://dx.doi.org/10.1007/s11207-013-0445-y}{Sol. Phys. 289, 2207 (2014)};

\bibitem{Grenier2015}
  Grenier, I. A., Black, J. H., Strong, A. W.,
  \href{https://dx.doi.org/10.1146/annurev-astro-082214-122457}{Annu. Rev. Astron. Astrophys. 53, 199 (2015)}
  
\bibitem{Tomassetti2015TwoHalo}
  Tomassetti, N.,
  \href{http://dx.doi.org/10.1103/PhysRevD.92.081301}{\PRD{} 92, 081301 (2015)}; 
  Tomassetti, N.
  \href{http://dx.doi.org//10.1088/2041-8205/752/1/L13}{\ApJ{} Lett. 752, L13 (2012)};
  
\bibitem{Feng2016}
  Feng, J., Tomassetti, N., Oliva, A.,
  \href{http://dx.doi.org/10.1103/PhysRevD.94.123007}{\PRD{} 94, 123007 (2016)}
  
\bibitem{Tomassetti2018PHeVSTime}
  Tomassetti, N., Bertucci, B., Bar\~{a}o, F., \etal,
  \href{http://dx.doi.org/10.1103/PhysRevLett.121.251104}{\Journal{\PRL}{121}{251104}{2018}}
   
\bibitem{Aguilar2015Proton} 
  Aguilar, M., \etal, 
  \href{http://dx.doi.org/10.1103/PhysRevLett.114.171103}{\Journal{\PRL}{114}{171103}{2015}}

\bibitem{Aguilar2015Helium}
  Aguilar, M., \etal, 
  \href{http://dx.doi.org/10.1103/PhysRevLett.115.211101}{\Journal{\PRL}{115}{211101}{2015}}
  
\bibitem{Boschini2017}
  Boschini, M. J., \etal, 
  \href{http://dx.doi.org/10.3847/1538-4357/aa6e4f}{\ApJ{} 840, 115 (2017)}
  
\bibitem{Corti2019}
  Corti, C., \etal, 
  \href{http://dx.doi.org/10.3847/1538-4357/aafac4}{\Journal{\ApJ}{871}{253}{2019}}
  
\bibitem{Corti2016} 
  Corti, C., Bindi, V., Consolandi, C., Whitman, K., 
  \href{http://dx.doi.org/10.3847/0004-637X/829/1/8}{\Journal{\ApJ}{829}{8}{2016}} 
  
\bibitem{Tomassetti2017TimeLag}
  Tomassetti, N., Orcinha, M., Bar\~{a}o, F., Bertucci, B., 
  \href{http://dx.doi.org/10.3847/2041-8213/aa9373}{\ApJ{} Lett. 849, 32 (2017)}
  
\bibitem{Tomassetti2015PHeAnomaly}
  Tomassetti, N.,
  \href{http://dx.doi.org/10.1088/2041-8205/815/1/L1}{\ApJ{} Lett. 815, L1 (2015)};
  
\bibitem{Tomassetti2017Universality}
  Tomassetti, N.,
  \href{http://dx.doi.org/10.1016/j.asr.2016.10.024}{\ASR{} 60, 815-825 (2017)}
  
\bibitem{Sun2015}
  Sun, X.,  Hoeksema, J. T., Liu, Y., and Zhao, J.,
  \href{10.1088/0004-637X/798/2/114}{\ApJ{} 798, 114 (2015)}
  
\bibitem{Usoskin2019} 
  Vaisanen, P., Usoskin, I., Mursula, K.,
  \href{https://doi.org/10.1029/2018JA026135}{J. Geophys. Res.: Space Phys. 124, 804-811 (2019)}
  
\bibitem{Borovsky2012}
  Borovsky, J. E.,
  \href{https://doi.org/10.1029/2011JA017499}{J. Geophys. Res. 117, A05104 (2012)}
  
\bibitem{Horbury2005}
  Horbury, T. S., Forman, M. A., \& Oughton, S.,
  \href{https://doi.org/10.1088/0741-3335/47/12B/S52}{Plasma Phys. Contr. Fusion, 47, B703-B717 (2005)} 
  
\end{thebibliography}
\end{document}